\documentclass[prb,twocolumn,showpacs,floatfix,superscriptaddress,amsmath,amssymb]{revtex4}
\usepackage{graphicx}
\usepackage{graphics}
\usepackage{color}
\usepackage{pst-all}

\newcommand{\beq}{\begin{equation}}
\newcommand{\eeq}{\end{equation}}
\newcommand{\beqa}{\begin{eqnarray}}
\newcommand{\eeqa}{\end{eqnarray}}
\newcommand{\nn}{\nonumber}

\newcommand{\s}{\textbf{S}}
\newcommand{\de}{\delta}

\begin{document}

\title{Spin-Peierls-like phases in magnetoelastic $J_1-J_2$ antiferromagnetic chain at 1/3 magnetization}

\author{ H.D.\ Rosales}
\affiliation{Departamento de F\'{\i}sica, Universidad Nacional de La Plata, C.C.\ 67, 1900 La
Plata, Argentina.}

\author{ G.L.\ Rossini}
\affiliation{Departamento de F\'{\i}sica, Universidad Nacional de La Plata, C.C.\ 67, 1900 La
Plata, Argentina.}

\date{\today}

\begin{abstract}
We investigate elastic deformations of spin
$S=1/2$ antiferromagnetic $J_1-J_2$ Heisenberg chains, at $M=1/3$
magnetization, coupled to phonons in the adiabatic approximation.
Using a bosonization approach we predict the existence of
non-homogeneous trimerized magnetoelastic phases. A rich ground
state phase diagram is found, including classical and quantum
plateau states for the magnetic sector as well as inequivalent
lattice deformations within each magnetic phase. The analytical
results are supported by exact diagonalization of small clusters.
\end{abstract}
\pacs{
75.10.Jm, 
73.43.Nq, 
75.30.-m  
}
\maketitle

\section{Introduction}
Frustrated spin systems have been continuously explored in the
last years. Frustration is considered a
key ingredient to induce unconventional magnetic orders or even
disorder, including spin-liquid states and exotic excitations. In
one-dimensional and quasi-one-dimensional models, quantum
antiferromagnets show many fascinating magnetic properties at low
temperatures which continue to attract an intense theoretical
activity. As representative of geometrically frustrated
homogeneous spin chains, one can consider the antiferromagnetic
spin $S=1/2$ zig-zag chain (for which compounds such as $CuGeO_3$
\cite{CuGeO3}, $LiV_2O_5$ \cite{LiV2O5} or $SrCuO_2$ \cite{SrCuO2}
are almost ideal prototypes) and three-leg antiferromagnetic spin
tubes (realized in $[(CuCl_2tachH)_3Cl]Cl_2$ \cite{spin-tube}). The chemistry of these compounds enables
the synthesis of single crystals much larger than the previously
observed organic analogs and, consequently, the achievement of
new and more precise experimental studies.

In this context, both experimental and theoretical interest on magnetoelastic chains was triggered
by the discovery of the spin-Peierls transition in $CuGeO_3$ \cite{CuGeO3(SP)} at zero
magnetization. This transition is an instability due to magnetoelastic effects which is
characterized (below  a critical temperature $T_{SP}$) by the opening of a spin gap and the
appearance of a dimerized lattice distortion at $M=0$, with the consequent modulation in spin
exchanges. Thus two related issues play together: the lattice distortion represents a cost in
elastic energy, while the spin exchange modulation modifies the magnetic spectrum.

A similar phenomenon can be analyzed in magnetoelastic systems at non zero magnetization, appearing
as most interesting the systems exhibiting magnetization plateaux. Moreover, it has been shown in
Ref.\ [\onlinecite{Vekua2006}] that a spin-phonon interaction in zig-zag chains explains a spin gap
opening as well as the presence of non-zero magnetization plateaux at low frustration, where they
are indeed absent in the case of non-elastic chains. Such plateaux are due to a mechanism of
commensurability between lattice distortions and spin modulation.

Regarding non-elastic zig-zag chains at $M=1/3$ magnetization plateaux, it was recently shown
\cite{Hida2005,Rosales2007} that small modulations of exchange couplings with period three on top
of a homogeneous zig-zag chain can drive a magnetic transition from a three-fold degenerate ground state
\cite{Okunishi2003} to either the so called classical plateau state ($CP$, where the spin
configuration resembles an Ising up-up-down state $\uparrow\uparrow\downarrow$) or the quantum
plateau state ($QP$, where the spin configuration resembles a quantum singlet-up state
$\bullet\bullet\!\!\uparrow$). Experimental and numerical evidence for a quantum plateau at $M=1/3$
was recently presented \cite{CuPOOH2006} for  $Cu_3(P_2 O_6 OH)_2$, a newly synthesized compound that
is very well described by spin $S=1/2$ antiferromagnetic chains
with period three modulated exchange couplings.

\begin{figure}[htbp]
   \centering
 \includegraphics[width=0.4\textwidth]{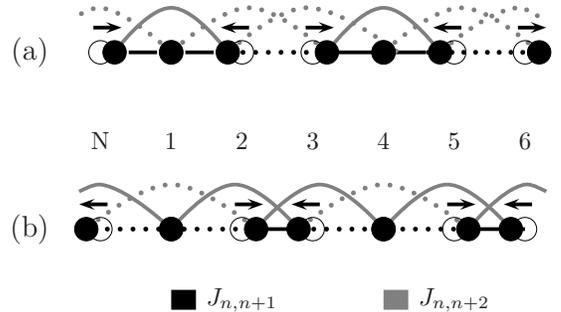}
    \caption{The upper panel describes a lattice deformation that enhances trimers;
    the lower panel corresponds to dimer enhancement.}
    \label{chains-introduction}
\end{figure}

Some insight about the magnetoelastic ground state can be obtained
from the mentioned fixed modulation results at $M=1/3$.
When one considers a $J_1-J_2$ chain with only nearest neighbors spin-phonon coupling,
a lattice deformation that brings closer two neighbors to the same site
(see Fig.\ \ref{chains-introduction}, upper panel)
induces a spin exchange modulation in $J_1$
forming open trimers. The ground state of the isolated trimer with $S_z=1/2$ indicates \cite{Hida2005}
the pinning of one of the classical plateau states, namely that with $\uparrow\downarrow\uparrow$
order on each trimer.
Instead, if two of every three sites group together forming dimers
(see Fig.\ \ref{chains-introduction}, lower panel),
the chain is driven to the quantum plateau state, that with spin singlets at each dimer.
A different situation arises when one considers also next-nearest neighbors
spin-phonon coupling, leading to modulations of both $J_1$ and $J_2$ exchanges.
The particular modulation discussed
in Ref.\ [\onlinecite{Rosales2007}] could be obtained (see again Fig.\ \ref{chains-introduction}, upper panel)
when $J_1$ and $J_2$ are modified so as to form closed trimers.
In contrast with the previous example, we have shown that in this case trimer enhancement drives this system
towards a quantum plateau state.
A natural question is then which magnetic configuration corresponds to a given lattice
deformation in the general case.

Motivated by the preceding discussion, we investigate in this paper
the possibility, suggested by the present authors and collaborators in Ref.\ [\onlinecite{Rosales2007}],
of a spin-Peierls like displacive transition
in an antiferromagnetic $S=1/2$ magnetoelastic
$J_1-J_2$ Heisenberg chain, when magnetization is set to M=1/3 by an external magnetic field.
We explore such a system with both nearest neighbors (NN) and next-nearest neighbors (NNN) spin-phonon
couplings in the adiabatic approximation, allowing for modulations of $J_1$ and $J_2$ exchanges.
This approach follows the recent discussion in Ref.\ [\onlinecite{Becca2003}],
where antiferromagnetic zig-zag spin chain compounds such as $CuGeO_3$ and $LiV_2O_5$ are argued to
present NNN  spin-phonon interactions at least of
the same order as the NN ones; indeed, a numerical study of such magnetoelastic zig-zag chains
at zero magnetization has lead to novel tetramerized spin-Peierls like phases.

We show that the magnetoelastic ground state at zero temperature indeed favors several period three
distortion patterns, stemming from a competition between elastic energy loss and magnetic energy
change. These patterns spontaneously break translation symmetry, with different phases depending on
the frustration ratio $J_2/J_1$ and the value of spin-phonon couplings. As we discuss below, there
are essentially four different situations that arise when a lattice deformation of period three
generates a spin exchange modulation at $M=1/3$: the lattice shows two kinds of period three
deformation patterns, namely (i) one tending to group three consecutive lattice sites into trimers
and (ii) another one tending  to group two of every three sites into dimers (see Fig.\
\ref{chains-introduction}). For each deformation pattern, depending on the microscopic parameters,
the spin sector adopts either (a) a classical plateau configuration, pinned in the lattice with the
$\downarrow$ spins sitting in the most convenient sites, or (b) a quantum plateau state, with the
spin singlets located at some convenient links. A rich phase diagram is built, including all of the
combinations of dimer-like and trimer-like deformations with both classical and quantum plateau
states.

The paper is organized as follows.
In  Section II  we present the model and its analytical treatment.
The spin sector is described within the bosonization approach,
while the phonon sector is described in the adiabatic approximation by classical static deformations.
In Section III we analyze this effective description by
considering all of the relevant perturbation terms as semiclassical potentials,
and draw a qualitative phase diagram with our results.
Special emphasis is put on the characterization of the ground state
phases that result from the combination of frustration and magnetoelastic effects
in different parameter ranges.
In Section IV we present the results of Lanczos exact diagonalization of small systems,
supporting the bosonization results.
Finally, in Section V we present a summary and conclusions of the present work.

\section{DESCRIPTION OF THE MODEL AND BOSONIZATION APPROACH}


We consider the lattice Hamiltonian of a frustrated spin $S=1/2$  Heisenberg chain,
which can be written as
\begin{eqnarray}
H_{M}&=&\sum_n\,\Big(J_{n,n+1}\,\s_n.\s_{n+1}+\,J_{n,n+2}\,\s_n.\s_{n+2}
\Big),
\end{eqnarray}
where $\s_n$ are spin operators at site $n$ and $J_{n,n+a}>0$ are antiferromagnetic
NN ($a=1$) and NNN ($a=2$) spin exchange couplings.
An uniform external magnetic field is also coupled to the spins in order to produce a global
magnetization $M=1/3$ ($M=1$ corresponding to saturation).

The interaction of spins in a homogeneous zig-zag chain ($J_{n,n+1}=J_1$, $J_{n,n+2}=J_2$)
with phonons is usually modeled by a linear expansion of the exchange couplings
around the non distorted values $J_1$ and $J_2$
\beqa
J_{n,n+1} &\approx& J_1(1-A(u_{n+1}-u_n)),\nn\\
J_{n,n+2} &\approx& J_2(1-B(u_{n+2}-u_n)),
\label{sph}
\eeqa
where $u_{n}$ is a scalar relevant coordinate for the displacement of ion $n$ from
its equilibrium position, and $A$, $B$ are called the spin-phonon couplings at NN and NNN sites.
The total Hamiltonian, including the elastic energy in the adiabatic approximation, is written as
\begin{eqnarray}
H_{T}&=&\frac{1}{2}K\sum_n(u_{n+1}-u_n)^2+\nn\\
&&+\sum_n\,\Big\{J_1\,\s_n.\s_{n+1}+J_2\,\s_n.\s_{n+2}\Big\}- \nn\\
&&-\sum_n\Big\{ J_1\,A(u_{n+1}-u_n)\s_n.\s_{n+1}+ \nn\\
&&+J_2\,B\,(u_{n+2}-u_n)\,\s_n.\s_{n+2}
\Big\},
\label{Htotal}
\end{eqnarray}
where $K$ is the homogeneous spring stiffness.
The first line corresponds to \emph{classical phonons} elastic energy ($H_{CP}$),
the second one to the homogeneous \emph{magnetic} Hamiltonian ($H_{M}$) and the rest to
the \emph{spin-phonon} interaction ($H_{I}$),
\beq
H_{T}=H_{CP}+H_{M}+H_{I}.
\label{Ht}
\eeq
Dimensionless parameters, convenient for numerical analysis, are used below.
They are introduced using $J_1$ as the energy scale
as follows:  $A \to \tilde{A}=(J_1/K)^{1/2} A$, $B \to \tilde{B}=(J_1/K)^{1/2} B$,
$u_n \to \tilde{u_n}=(K/J_1)^{1/2} u_n$ and $J_1,J_2 \to \alpha=J_2/J_1$.


    In order to observe semi-quantitatively the low energy properties of the model given by Eq.(\ref{Htotal}),
we employ the bosonization method which is generally powerful for the description of
one-dimensional spin chains (see for instance Ref.\ [\onlinecite{Giamarchi2004}]).

    We start with the homogeneous magnetic Hamiltonian $H_{M}$.
To obtain the corresponding low-energy theory
one first applies the exact Jordan-Wigner transformation mapping spins onto
spinless lattice fermions $\psi_n$,
then one introduces a continuum coordinate $x=n a$ with $a$ the lattice spacing and
writes a linear approximation for the low energy degrees of freedom around the Fermi level
in terms of left and right-moving continuum fermions;
the Fermi wave-vector $k_F$ depends on the magnetization.
For $M=1/3$ one gets $k_F=\pi /3a$,  then
\begin{equation}
\label{rules-low-energy}
 \psi_n \approx e^{i\pi n/3}\psi_R(n a)+e^{-i\pi n/3}\psi_L(n a).
\end{equation}
The continuum fermions are spinless and massless, allowing for Abelian bosonization;
the complete Hamiltonian is finally mapped into a Gaussian term
\beq
\frac{v}{2}\int dx \big[\frac{1}{K_L}(\partial\phi)^2+K_L\,(\partial\tilde{\phi})^2\big]
\label{Gaussian}
\eeq
plus several vertex operators that are kept only when they are commensurate (non-oscillating in space)
and constitute relevant perturbations to the Gaussian conformal field theory.
Here $\phi$ is a compactified boson field  defined on a circle,
$\phi \equiv \phi+ \sqrt{\pi}$, and $\tilde\phi$ is its dual
field defined by $\partial_x\tilde\phi = \partial_t\phi$. The
parameters $v$ and $K_L$ (Fermi velocity and Tomonaga-Luttinger
parameter respectively) depend on the microscopic parameters of the lattice
Hamiltonian $H_M$; $v$ is proportional to $a J_1$, while $K_L$ is dimensionless.

    A particular feature of the $M=1/3$ situation is that $k_F=\pi /3a$
makes commensurate a triple Umklapp process \cite{Hida2005},
providing a perturbation term of the form
\beq
-\frac{ g_3\,v}{2\pi^2a^2}\int dx\cos(3\sqrt{4\pi}\phi)
\label{cos3phi}
\eeq
in $H_M$. The coefficient $g_3$ is non-universal and, as well as $v$ and $K_L$, depends on the
renormalization group procedure.

    It is known numerically \cite{Okunishi2003,Okunishi2004} that the homogeneous magnetic Hamiltonian
$H_M$ describes a gapless Tomonaga-Luttinger (TL) phase
for $0 < J_2/J_1 < \alpha_c=0.56$
\footnote{The critical value for $J_2/J_1$ has also been reported as 0.487 using level spectroscopy, see Ref.\
[\onlinecite{Okunishi2004}].}.
For $\alpha_c < J_2 / J_1 \lesssim 1.25$ there exists a strong magnetization plateau at
$M=1/3$.
Comparison of bosonization with these results shows that the
the Tomonaga-Luttinger parameter should be $K_L > 2/9$ for $J_2/J_1 \lesssim \alpha_c$,
as this renders the perturbation in Eq.\ (\ref{cos3phi}) irrelevant.
Then the coefficient $g_3$ flows to zero under the renormalization group and
the effective theory describes a gapless TL phase.
On the other hand, for $ \alpha_c \lesssim J_2/J_1 \lesssim 1.25$,
it should be $K_L < 2/9$, making Eq.\ (\ref{cos3phi}) a relevant perturbation.
Thus this term opens a magnetic gap and
explains the magnetization plateau \cite{Lecheminant2004} observed in this range.
Moreover, the plateau ground state is known to be three-fold degenerate,
with translation symmetry spontaneously broken to an up-up-down
configuration \cite{Okunishi2003}; such configurations are described
by the pinning of the bosonic field in one of the minima of Eq.\ (\ref{cos3phi})
considered as a semiclassical potential energy \cite{Lecheminant2004},
provided that $g_3 > 0$.
We will in consequence qualitatively represent the plateau by
the behavior of the non-universal coefficient $g_3 \geq 0$
as being smooth and non-vanishing only for $ \alpha_c \lesssim J_2/J_1 \lesssim 1.25$,
with a maximum at some intermediate value of $J_2/J_1$.
We will not study here the regime  $J_2/ J_1 > 1.25$, where the $M=1/3$ plateau is not present;
this should be better done by starting with two spin chains with strong exchange $J_2$,
weakly coupled by a zig-zag interaction $J_1$.


Next, we consider the lattice deformations.
From the knowledge of the $M=1/3$ plateau magnetic ground state in the homogeneous $J_1-J_2$ chain with
$J_2/J_1 > \alpha_c$, one can argue that an adiabatic lattice deformation caused by the spin-phonon coupling
in Eq.\ (\ref{Htotal}) will have period three.
This is also supported by bosonization, as such a deformation is commensurate with $k_F=\pi/3a$.
Moreover, as discussed in Ref.\ [\onlinecite{Vekua2006}],
even for $J_2/J_1 < \alpha_c$ period three deformations
cause commensurability of relevant perturbations at $M=1/3$
and provide a mechanism for a spin gap (magnetization plateau)
in this regime.
Numerical evidence of the dominance of period three lattice deformations,
obtained from self consistent computations, was also given in [\onlinecite{Vekua2006}].
A uniform deformation, leading to global size change, can also appear \cite{Becca2003};
this would produce a uniform shift in $J_1$ and $J_2$, which is unessential
to our present analysis.

We will consider in this paper the most general period three deformation,
without collective displacement, given by
\beq
u_n=\frac{u_0}{\sqrt{3}}\,\sin(\frac{2\pi}{3}\,n - \chi),
\label{uns}
\eeq
with the amplitude $u_0$ and a relative phase $\chi$ as free parameters.
Our purpose is to search for the deformation that minimizes the magnetoelastic energy.
Once a minimum of  the total energy is found,
the amplitude $u_0$ will indicate the deformation strength and
the phase $\chi$ will relate the deformation pattern to the corresponding spin
ground state characterized by the value of $\phi$ at the potential minimum.

From Eq.\ (\ref{uns}) the distortion of the NN bond length between sites $n$ and $n+1$,
denoted by $\de_n=u_{n+1}-u_{n}$,
is parameterized by
\begin{small}
\begin{equation}
\de_n=u_0\,\cos\big(\frac{2\pi}{3}\big(n+\frac{1}{2}\big) - \chi\big),
\label{deformation1}
\end{equation}
\end{small}
while the NNN distortion is given by
\begin{small}
\begin{equation}
\de_{n+1}+\de_n=u_0\,\cos\big(\frac{2\pi}{3}\big(n+1) - \chi\big).
\label{deformation2}
\end{equation}
\end{small}
%
The elastic energy cost associated to deformations in Eq.\ (\ref{uns}) reads simply
\beq
H_{CP}/J_1=\frac{1}{4}N\tilde{u}^2_0.
\label{H_CP}
\eeq

Finally, we consider the spin-phonon interaction Hamiltonian $H_I$ induced
by lattice deformations in Eq.\ (\ref{uns}).
Following the bosonization procedure one generates an extra renormalization of $v$ and $K_L$
and perturbation terms of the form
\beq
\frac{\tilde{u}_0\,v}{2\pi^2a^2}
\int dx \left(f_1\cos(\sqrt{4\pi}\phi + \chi)+f_2\cos(2\sqrt{4\pi}\phi - \chi)\right),
\label{cos1cos2}
\eeq
thus introducing first and second harmonics of the boson field
with coefficients proportional to the deformation amplitude $\tilde{u}_0$.
Notice that these operators
are more relevant than the third harmonic in Eq.\ (\ref{cos3phi}),
and should be kept as well in the $J_2/J_1 > \alpha_c$ regime as
for $J_2/J_1 < \alpha_c$, as far as $K_L <1/2$.
Even though the coefficients are non-universal and subject to renormalization,
it is useful to report that a first order perturbative computation yields
$f_1 \sim \tilde{A}(1-C_1\,J_2/J_1)$
and
$f_2 \sim -\tilde{A}(1+C_2\, q\,J_2/J_1)$,
where $q=\tilde{B}/\tilde{A}$ and $C_1,C_2$ are positive constants with $C_2 \ll C_1$.
For a qualitative description, we will assume that $f_1$ and $f_2$
depend on the microscopic parameters as
suggested by these bare expressions.


Putting all together, we can write the complete effective theory as
\beq
H_T=H_{CP}+H_{free}+V_{eff}
\eeq
where $H_{CP}$ is the classical elastic contribution given in Eq.\ (\ref{H_CP}),
\beq
H_{free}=\frac{v}{2}\int dx \big[\frac{1}{K_L}(\partial\phi)^2+K_L\,(\partial\tilde{\phi})^2\big]
\label{Hfree}
\eeq
is the Gaussian part of the compactified boson action and
\beqa
V_{eff}&=&\frac{v}{2\pi^2a^2}\int dx
\big[\tilde{u}_0\,f_1\cos(\sqrt{4\pi}\phi + \chi) \label{Veff}\\
&& +\tilde{u}_0\,f_2\cos(2\sqrt{4\pi}\phi - \chi)
-\,g_3 \cos(3\sqrt{4\pi}\phi)\big]
\nn
\eeqa
is the bosonic self-interaction potential defining a triple sine-Gordon theory
\cite{HoghJensen1982}.
Extensive analysis of competition between harmonics
in multi-frequency sine-Gordon theories has been
performed \cite{Delfino1997,Nersesyan2000,Takacs2001}, mainly focused on the double sine-Gordon
model. The three-frequency case has also been recently discussed \cite{Toth2004}.
For our purpose it will be enough to perform a semiclassical treatment, as detailed in the next section.

\section{Semiclassical analysis of the effective theory}

The aim of the present work is to search for the possibility of elastic deformations
that lower the magnetoelastic energy with respect to the homogeneous non-deformed case.
The simplest analysis of the effective theory obtained in the previous section,
which has proved to be useful in related cases
\cite{Lecheminant2004,Hida2005,Rosales2007},
consists in treating the self-interaction terms in Eq.\ (\ref{Veff}) as a
classical potential to be evaluated in constant field configurations.

Within this approximation the energy per site depends on three configuration parameters,
$\tilde{u}_0$, $\phi$ and $\chi$, and is readily evaluated to
\beqa
\lefteqn{\epsilon(\tilde{u}_0,\phi,\chi) \equiv \frac{E}{J_1\ N}=
\frac{1}{4}\tilde{u}_0^2 -
\frac{g_3}{2\pi^2}\cos(3\sqrt{4\pi}\phi)} \label{energy} \\
&&+ \frac{\tilde{u}_0}{2\pi^2}(f_1\cos(\sqrt{4\pi}\phi + \chi)
+f_2\cos(2\sqrt{4\pi}\phi - \chi)),
\nn
\eeqa
so that  the minima can be found analytically.
Notice that this expression is invariant under simultaneous shifts
$\sqrt{4\pi}\phi \to \sqrt{4\pi}\phi+2\pi/3$,
$\chi \to \chi - 2\pi/3$, in relation with the three equivalent locations of period three
structures on the chain.
This allows  to restrict the analysis to $0 < \sqrt{4\pi}\phi \leq 2\pi/3$ without loss of generality.
Also a shift $\chi \to \chi+\pi$ is equivalent to changing the sign of $u_0$, allowing to consider
$0 \leq \chi < \pi$. We report results within these restricted ranges.

Among several local minima of the potential, the semiclassical energy
is always found in one of the following situations:
\begin{enumerate}

\item $\sqrt{4\pi}\phi=2\pi/3$, $\chi = \pi/3$, $\tilde{u}_0 = (f_1 + f_2)/\pi^2$,
where the energy is evaluated to
\beq
\epsilon=-\frac{g_3}{2\pi^2}-\frac{(f_1 + f_2)^2}{4\pi^4}.
\label{solCP}
\eeq

\item $\sqrt{4\pi}\phi=\pi/3$, $\chi = 2\pi/3$, $\tilde{u}_0 = (f_1 - f_2)/\pi^2$,
where the energy is evaluated to
\beq
\epsilon=\frac{g_3}{2\pi^2}-\frac{(f_1 - f_2)^2}{4\pi^4}.
\label{solQP}
\eeq

\end{enumerate}

Before drawing a phase diagram, we discuss the physical content of the possible phases.
Following the usual bosonization rules to map $\phi$ to spin variables \cite{CabraPujol},
the value $\sqrt{4\pi}\phi=2\pi/3$ in the first solution indicates that the spin sector adopts a
classical plateau configuration $CP$ \cite{Hida2005},
which corresponds to selecting one of the $\uparrow\uparrow\downarrow$ degenerate ground states
of the homogeneous chain plateau. We will call its energy $\epsilon_{CP}$.
The relative phase $\chi = \pi/3$ plays together with the sign of $\tilde{u}_0$ in determining
the elastic deformation.
For $f_1 + f_2 > 0$ one finds a trimer-like elastic deformation
grouping blocks of three spins
($T$, see Fig.\ \ref{chains-introduction}, upper panel);
in the opposite case a dimer-like deformation is set, alternating two closer spins with a more separated one
($D$, see Fig.\ \ref{chains-introduction}, lower panel).
In the second solution,
the value $\sqrt{4\pi}\phi=\pi/3$ is not one of the minima
of the homogeneous chain potential, but it signals that the spin sector adopts
a state that enhances quantum singlets in a
$\bullet\bullet\!\uparrow$
quantum plateau configuration \cite{Hida2005} $QP$.
The corresponding energy will be called $\epsilon_{QP}$.
The relative phase $\chi = 2\pi/3$ in this solution indicates that
the lattice deformation is dimer-like ($D$) for $f_1 - f_2 > 0$
and trimer-like ($T$) otherwise.

Depending on the coefficients $f_1$, $f_2$ and $g_3$,
which in turn depend on the microscopic parameters,
one of these solutions is selected as the global minimum and
determines the magnetoelastic ground state phase.

\begin{figure}[htbp]
\centering
\includegraphics[width=8.5cm,height=5.5cm]{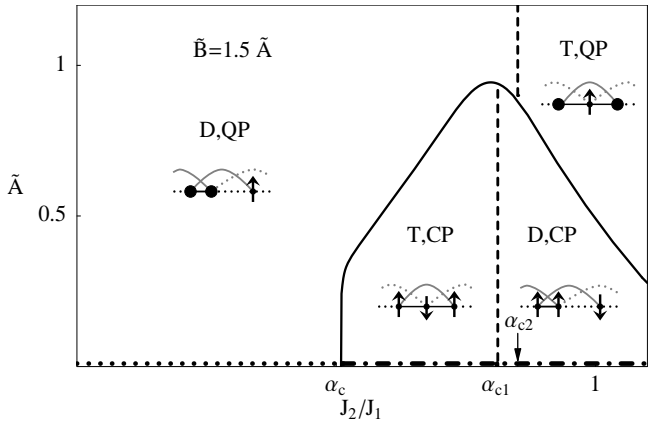}
\caption{
Schematic magnetoelastic phase diagram with NN and NNN spin-phonon couplings related by
$\tilde{B}=1.5 \,\tilde{A}$.
The classical up-up-down and quantum plateau
phases are labelled by $CP$ and $QP$ respectively.
The dimer and trimer elastic phases are labelled by $D$ and $T$.
Magnetoelastic patterns in each phase are shown by pictorial diagrams.
}
\label{phases}
\end{figure}

In order to present a schematic phase diagram,
we assume the qualitative phenomenological dependence of $f_1$, $f_2$ and $g_3$
on the microscopic parameters detailed in the previous section.
Following Ref.\ [\onlinecite{Becca2003}]
we have chosen a ratio $\tilde{B}=1.5 \,\tilde{A}$,
which is used in the rest of the paper,
as representative of materials where
the NNN spin-phonon coupling plays an important role.
The magnetoelastic phases found are shown in Fig.\ \ref{phases}.
We have checked that,
within our approximations, all phase transitions result from
level crossing of the above described local minima and can be then classified as first order.

\begin{figure}[htb]
\includegraphics[width=9.5cm,height=6.8cm]{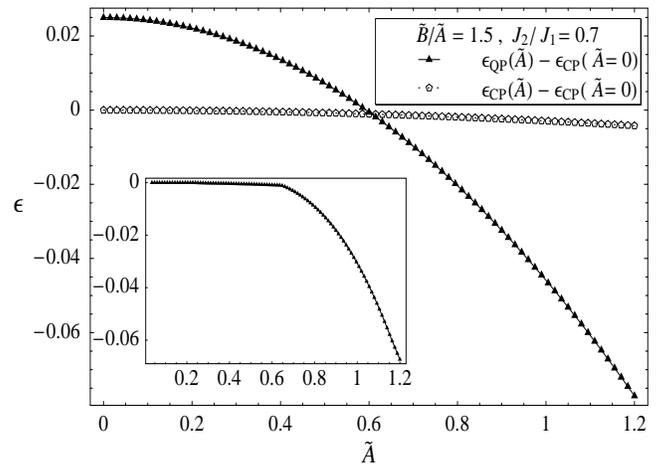}
\caption{
Semiclassical energies for the classical and quantum plateaux minima in terms of the spin-phonon coupling
$\tilde{A}$, for $\tilde{B}=1.5\,\tilde{A}$ and $J_2/J_1=0.7$.
The inset shows the ground state energy obtained by exact diagonalization of a system with $N=24$ sites,
after fitting the zero-energy level.}
\label{EvsAB15J07}
\end{figure}

For the sake of illustrating the analysis leading to Fig.\ \ref{phases},
we show in Fig.\ \ref{EvsAB15J07}
the evolution of the energies $\epsilon_{CP},\, \epsilon_{QP}$ as functions of
the NN spin-phonon couplings $\tilde{A}$, $\tilde{B}=1.5 \,\tilde{A}$,
fixing $J_2/J_1=0.7$. This situation lies well inside the homogeneous plateau regime, $J_2/J_1 > \alpha_c$.
The coefficients $f_1,\, f_2$ are
evaluated according to the first order bare result given in the previous section.
The level crossing at $\tilde{A}_c\simeq 0.65$ shows the transition from $CP$ magnetic phase to $QP$.
We remark that this transition is very different than that recently observed by the authors in
[\onlinecite{Rosales2007}], where the system passes from $CP$
to a $Z_2$ broken symmetry phase and only then to a $QP$ phase through an Ising
like second order transition.

Within each magnetic phase one can also identify the different elastic phases.
Given $\tilde{A}$, we find  critical values of $J_2/J_1 $
where $\tilde{u}_{0}$ changes sign:
in the $CP$ phase, the equation
$f_1(\alpha_{c1})=-f_2(\alpha_{c1})$ defines a critical line $J_2/J_1=\alpha_{c1}$
such that for $J_2/J_1 < \alpha_{c1}$ the system adopts a trimer like lattice distortion $T$
while for $J_2/J_1 > \alpha_{c1}$ the deformation is dimer like $D$.
In contrast, in the $QP$ phase we find a critical line $\alpha_{c2}$ where
$f_1(\alpha_{c2})=f_2(\alpha_{c2})$, being the lattice distortion
of type $D$ for $J_2/J_1 < \alpha_{c2}$ and of type $T$ for  $J_2/J_1 > \alpha_{c2}$.
Using the bare expressions for $f_1,f_2$,
these critical lines do not depend on $\tilde{A}$.

A similar analysis can be made for $J_2/J_1 \lesssim \alpha_c$.
The most important difference is that
in this region the Tomonaga-Luttinger parameter is
$K_L > 2/9$ and the third harmonic is irrelevant.
We represent this situation by setting $g_3=0$.
As mentioned before, the magnetization plateau at $M=1/3$ is induced
by the coupling to the lattice \cite{Vekua2006}
through the first and second harmonics in Eq.\ (\ref{Veff}).
Unlike the previous case, there is no level crossing
between $\epsilon_{QP}$ and $\epsilon_{CP}$;
the absolute energy minimum always corresponds to $\epsilon_{QP}$,
selecting the $QP$ magnetic phase.
We also find $(f_1-f_2)>0$ in the whole region,
so that the elastic phase is of type $D$.

The relative position of the elastic deformations and the magnetization profile
at each phase is determined by the corresponding values of $\phi$ and $\chi$, as
can be found using Eq.\ (\ref{uns}) and bosonization formulae.
The four phases are described by diagrams in Fig.\ \ref{phases}.

Different ratios $\tilde{B}/\tilde{A}$ can be analyzed similarly. We have observed that lowering
this ratio produces an increase in the region characterized by
the classical plateau and trimer-like deformations, with higher values of both
$\alpha_{c1}$ and $\tilde{A}_c$.

It is important to notice that the deformation amplitude $\tilde{u}_0$ is proportional to
$f_1$, $f_2$, which are in turn proportional
to the dimensionless spin-phonon couplings $\tilde{A}$, $\tilde{B}$.
In Fig.\ \ref{phases}, the elastic pattern evolves to the homogeneous limit
as $\tilde{A} \rightarrow 0$. By construction, the effective theory in Eqs.\
(\ref{Hfree}, \ref{Veff}) describes in this limit a Tomonaga-Luttinger phase for $J_2/J_1<\alpha_c$
and a gapped sine-Gordon phase with triple degenerate ground state for $J_2/J_1 > \alpha_c$.

It is also interesting to mention that if our approach remains valid
in the limit $J_2/J_1 \rightarrow 0$, describing a single NN spin chain,
the system adopts a dimer like elastic deformation which in turn induces a trimerized
modulation with one larger and two smaller NN spin exchanges (c.f.\ Fig.\ \ref{chains-introduction},
lower panel).
This model was recently studied by quantum Monte Carlo simulations of large systems
in connection with $Cu_3(P_2 O_6 OH)_2$ \cite{CuPOOH2006}, finding exactly the quantum plateau
magnetic phase predicted by our analysis.

\section{Numerical analysis}

In order to support the bosonization results in the previous section,
we performed a numerical analysis of the Hamiltonian in Eq.\ (\ref{Htotal})
by exact diagonalization of small clusters of size up to $N=24$ sites
with periodic boundary conditions.

The strategy is the following: period three elastic deformations without collective displacement
are parameterized by two independent bond distortions, say $\delta_1=u_2-u_1$ and $\delta_2=u_3-u_2$,
while $\delta_3=-\delta_1-\delta_2$ and $\delta_{n+3}=\delta_n$.
For given values of $J_2/J_1$ and $\tilde{A}$, fixing $\tilde{B}/\tilde{A}$ and $M=1/3$,
we computed by Lanczos diagonalization \cite{Lanczos}
the exact ground state energy of the total Hamiltonian
in Eq.\ (\ref{Htotal})
for a wide range of elastic deformations $(\delta_1, \delta_2,\delta_3)$
and then selected the absolute minimum.

We found that, in accordance with bosonization results,
the lowest energy configuration is always obtained (except for equivalent lattice translations)
at one of lattice distortions patterns shown in Fig.\ \ref{chains-introduction}:
\begin{enumerate}
\item
$(\delta_1, \delta_2, \delta_3)=(-\frac{1}{2}\Delta, -\frac{1}{2}\Delta, \Delta)$
that corresponds to the trimer-like phase ($T$), or
\item
$(\delta_1, \delta_2, \delta_3)= (\frac{1}{2}\Delta, \frac{1}{2}\Delta, -\Delta)$
that corresponds to the dimer-like ($D$).
\end{enumerate}

In order to characterize the magnetic phases, we also computed the local magnetization profile $<S^z_n>$
for the ground state. The order parameter
\begin{eqnarray}
M_S&=&\frac{1}{N}\sum_n\cos(\frac{2\pi}{3}(n-2))<S^z_n>
\label{MS}
\end{eqnarray}
introduced in Ref.\ [\onlinecite{Rosales2007}], which is positive for the quantum plateau (QP) configuration
and negative for the classical plateau (CP), is used to report the results.
We found that both magnetic phases are realized at some region of either the $T$ or the $D$ elastic phases.

\begin{figure}[htbp]
\centering
\includegraphics[width=8.7cm,height=5.9cm]{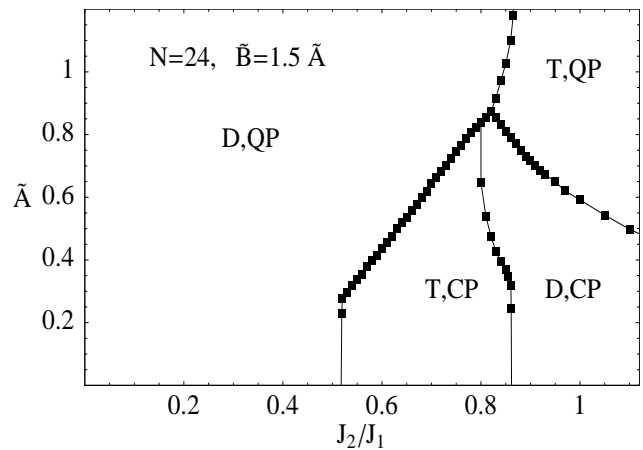}
\caption{Magnetoelastic phase diagram obtained by exact diagonalization of $N=24$ sites.
Spin-phonon couplings ratio is set to $\tilde{B}/\tilde{A}=1.5$.}
\label{phasenumB15}
\end{figure}

\begin{figure}[htbp]
    \centering
        \includegraphics[width=0.42\textwidth]{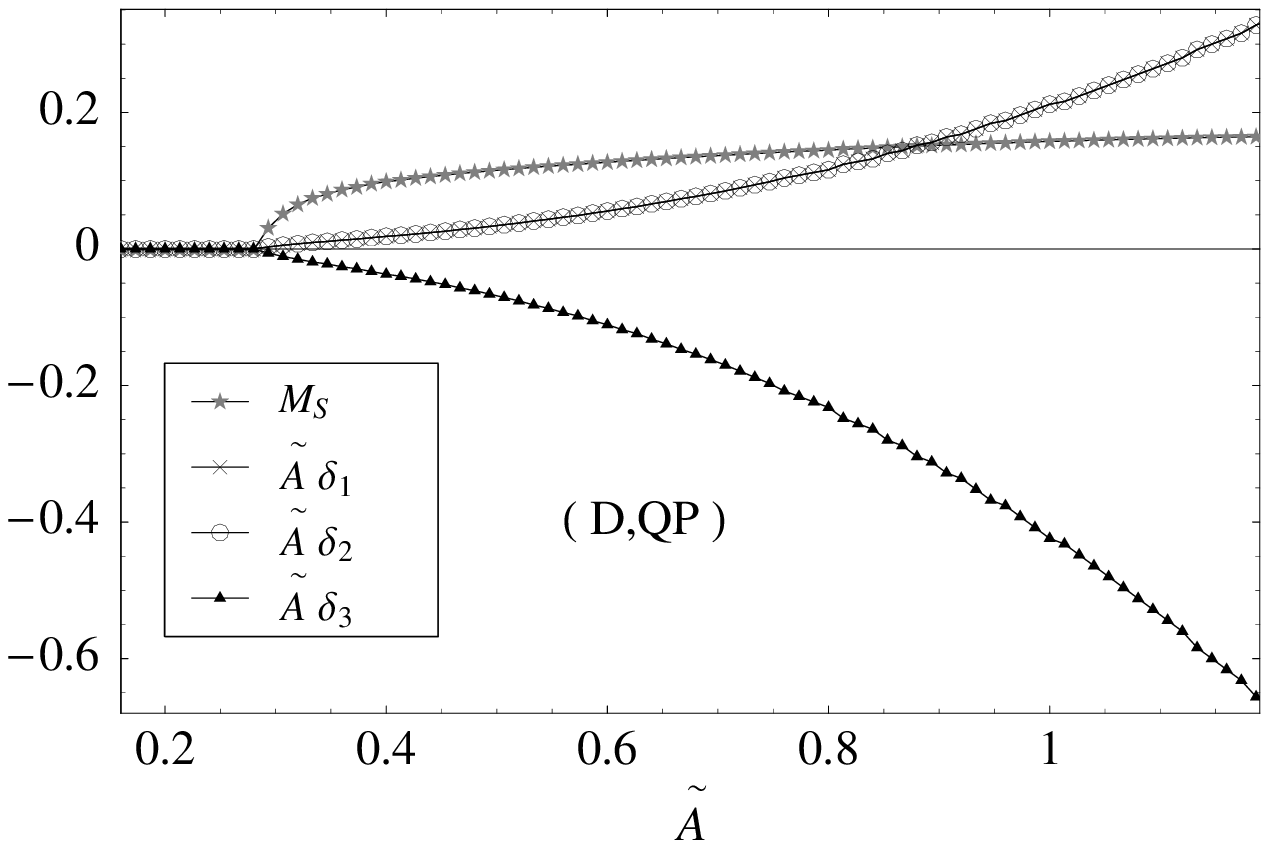}
        \includegraphics[width=0.42\textwidth]{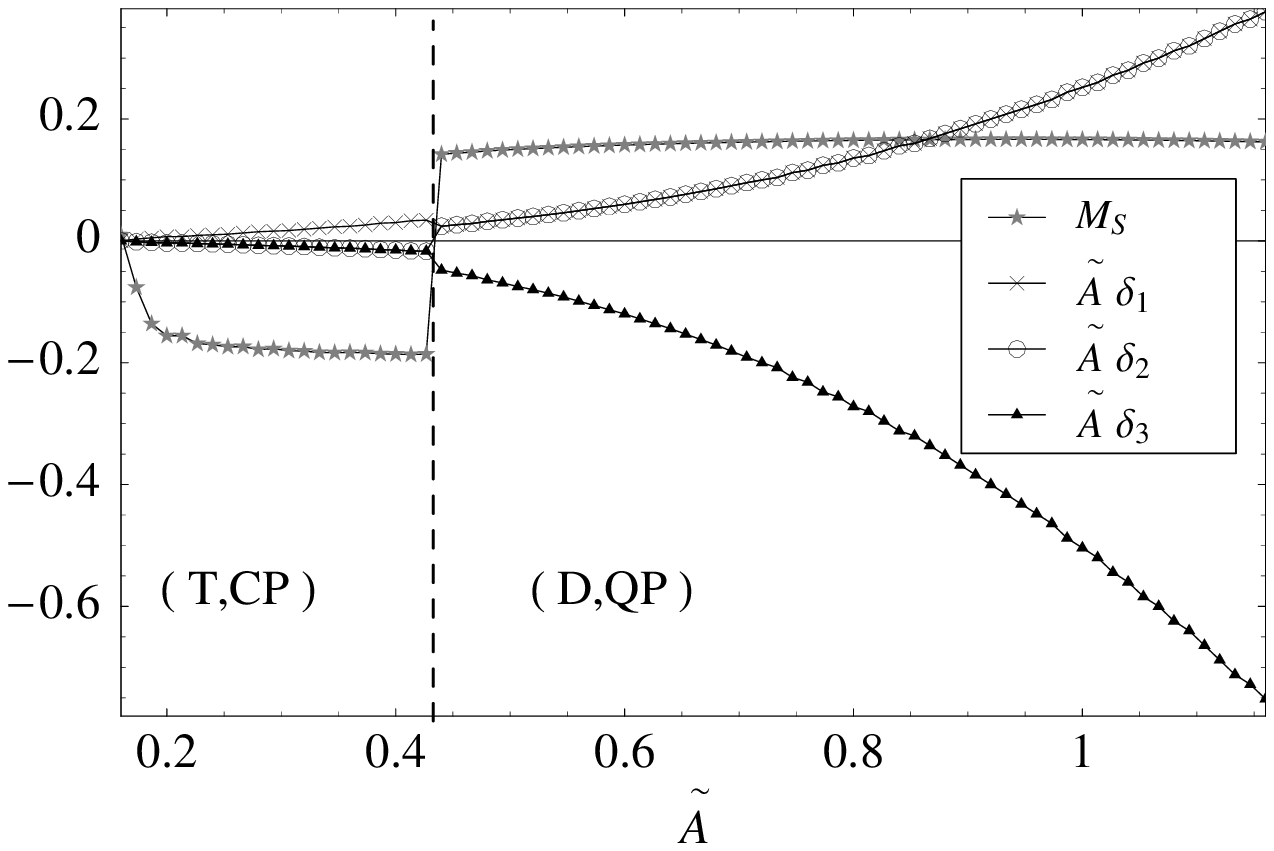}
        \includegraphics[width=0.42\textwidth]{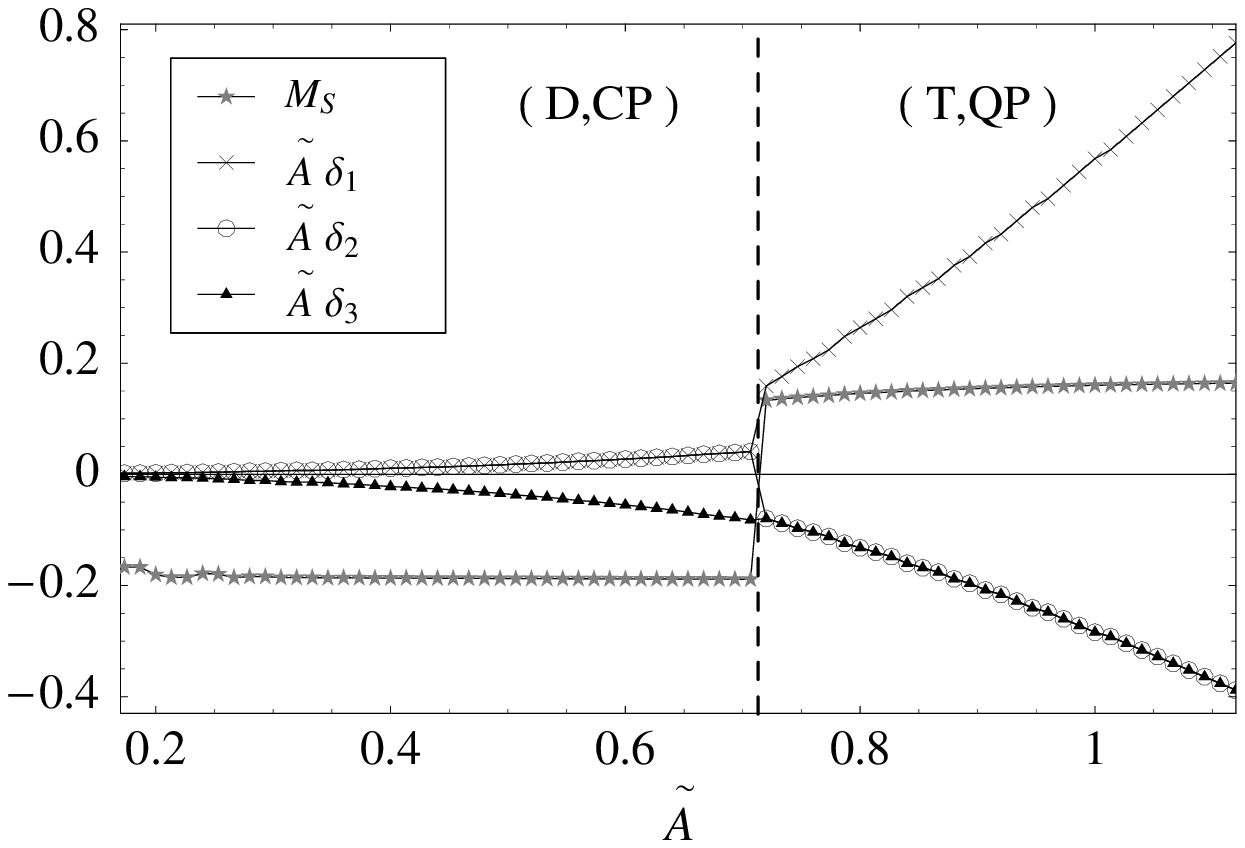}
\caption{Lattice distortions $\delta_{1,2,3}$ and order parameter $M_S$ as a function of
$\tilde{A}$ for $\tilde{B}/\tilde{A}=1.5$ and $J_2/J_1=0.3$ (upper panel), $0.6$ (middle panel),
$0.9$ (lower panel). Each panel corresponds to a vertical scan in Fig.\ \ref{phasenumB15}. Small
$\tilde{A}$ is not accesible due to finite size effects.} \label{Msd1d2d3B15}
\end{figure}

A thorough scanning of the
$\tilde{A} > 0, J_2/J_1 > 0$
plane was made for $N=24$ sites, keeping $\tilde{B}=1.5\,\tilde{A}$.
The magnetoelastic phases found are shown in Fig.\ \ref{phasenumB15}.
Representative scans at $J_2/J_1= 0.3,\,  0.6,\, 0.9$ are shown in Fig.\ \ref{Msd1d2d3B15},
showing $\delta_1$, $\delta_2$, $\delta_3$ and $M_S$ as functions of $\tilde{A}$.
Notice that the deformation amplitude decreases for small $\tilde{A}$ (a limit that corresponds to
large stiffness $K$);
due to finite size effects \cite{Feiguin1997} there is no deformation below some finite value of
$\tilde{A}$, which is seen to diminish with size by comparing $N=12,18,24$ sites.
The $J_2/J_1=0.3$
scan shows the phase $D,\,QP$ for all $\tilde{A}$. In the $J_2/J_1=0.6$ case one can clearly
observe the first order transition at some value of $\tilde{A}$ (which in general depends on
$J_2/J_1$), with finite jumps both in the deformations and the magnetic order parameter, from the
$T,\, CP$ to the $D,\, QP$ phase. The same happens in the $J_2/J_1=0.9$ case, with a transition
from the $D,\, CP$ to the $T,\, QP$ phase. The region $0.8 < J_2/J_1 < 0.9$ shows that the critical
line for transition between $T$ and $D$ phases slightly depends on $J_2/J_1$; comparison with Fig.\
\ref{phases} indicates that renormalization effects on the bare coefficients $f_1, \, f_2$ are not
strong enough to impede our qualitative bosonization analysis.

We have also analyzed different spin-phonon couplings, confirming the bosonization prediction
that lowering the ratio $\tilde{B}/\tilde{A}$
produces an increase in the region characterized by the classical plateau and trimer-like
deformations ({\em c.f.} Fig.\ \ref{phases}), with higher values of both $\alpha_{c1}$ and $\tilde{A}_c$ .

In summary, the numerical results confirm the semiclassical analysis
given in previous section.

\section{CONCLUSIONS}

In the present work we have shown the existence of different magnetoelastic phases
in $S=1/2$ zig-zag antiferromagnetic $J_1-J_2$ chains
coupled to adiabatic phonons through nearest and next-nearest neighbor spin exchanges,
at $M=1/3$ magnetization.
At zero temperature this situation corresponds to a magnetization plateau,
either existing for the non-distorted homogeneous chain
with high enough frustration \cite{Okunishi2003,Lecheminant2004}
or induced by spin-phonon coupling at lower frustration \cite{Vekua2006}.

We performed a semiclassical analysis of the bosonized effective theory,
supported by numerical exact diagonalization of small clusters up to 24 spins.
We found that several spin-Peierls like phases describe the ground state of the system,
depending on the microscopic parameters $J_2/J_1$ and spin-phonon couplings.
In each of these phases a non trivial elastic deformation is favored,
grouping together blocks of two or three spins,
while the magnetic sector adopts classical or quantum plateau states.

A detailed analysis of a particular case,
chosen as representative of materials with large ratio of
next-nearest to nearest neighbors spin-phonon couplings \cite{Becca2003},
shows the following magnetoelastic phases at zero temperature:

\noindent (i) an up-up-down magnetic phase with a trimer-like lattice distortion
when frustration is just enough to produce
the $M=1/3$ magnetization plateau in the homogeneous chain
and spin-phonon couplings are low.

\noindent (ii) an up-up-down magnetic phase with a dimer-like lattice distortion
for low spin-phonon couplings and higher frustration.

\noindent (iii) a quantum plateau magnetic phase with dimer-like lattice distortion
for large spin-phonon couplings and low frustration. This phase is present
even for  such low frustrations
that would not produce a magnetization plateau in absence of spin-phonon interaction.

\noindent (iv) a quantum plateau magnetic phase with trimer-like lattice distortion
for large spin-phonon couplings and high frustration.

Once the existence of non trivial magnetoelastic phases at zero temperature is proved,
a natural question is to analyze the possibility of a
spin-Peierls like transition  in three dimensional materials with
quasi-one-dimensional magnetic structure.
Since a high temperature phase is expected
to recover translation invariance,
such a transition could take place at some finite temperature
\cite{Dobry2007}
, while an external magnetic field
maintains the magnetization $M=1/3$.
However, a finite temperature study should also take into account the
eventual smoothing of the magnetization plateau.
The critical temperature
and the behavior of thermodynamic functions at the conjectured
transition is suggested for future investigation.

{\em Acknowledgments:} the authors thank T.\ Vekua for bringing attention into the subject,
A.O.\ Dobry and D.C.\ Cabra for valuable comments and M.D.\ Grynberg for numerical assistance in
Lanczos diagonalization. This work was partially supported by CONICET (Argentina), PICT ANCYPT
(Grant 20350), and PIP CONICET (Grant 5037).


\end{document}